\documentclass[twocolumn,showpacs,preprintnumbers,amsmath,amssymb]{revtex4}

\usepackage{graphicx}
\usepackage{dcolumn}
\usepackage{bm}
\newcommand{\Obs}[2]{\textbf{Observation #1.} \textit{#2}}
\newcommand{\Proof}{\textbf{Proof.} }

\newcommand{\ket}[1]{\big|#1\big>}
\newcommand{\bra}[1]{\big<#1\big|}

\newcommand{\proj}[1]{\ket{#1}\bra{#1}}
\newcommand{\p}[0]{^{(\pm)}}
\newcommand{\m}[0]{^{(\mp)}}
\newcommand{\ot}[0]{\otimes}

\begin{document}

\preprint{APS/123-QED}

\title{Generalised Smolin states and their properties}

\author{Remigiusz Augusiak}
\author{Pawel Horodecki}
\affiliation{Faculty of Applied Physics and Mathematics \\
Gda\'nsk University of Technology, Gda\'nsk, Poland}

\date{\today}

\begin{abstract}
Four qubit bound entangled Smolin states are generalised in a natural way to 
even number of qubits. They are shown to maximally violate simple
correlation Bell inequalities and, as such, to reduce communication 
complexity, though they do not admit quantum security.
They are also shown to serve for remote quantum information concentration 
as like in the case of the original four qubit states. 
Application of the information concentration to the process of  
unlocking of classical correlations and quantum entanglement 
by quantum bit is pointed out.  
\end{abstract}

\pacs{Valid PACS appear here}
\maketitle
\section{Introduction}
\label{I}

Quantum entanglement \cite{EPR},  \cite{Sch} is a very important resource in
quantum information theory (QIT) \cite{Ksiazki}. It contributes to fundamental
quantum information phenomena \cite{Artur,geste,teleportacja,Shor} and represents itself the quality that is not present in classical world. Entanglement 
of pure states have been shown to be incompatible with any local hidden
models since it violates well-known Bell inequalities
\cite{Bell}. It has been also proved to be an optimal resource for quantum 
information. The case of mixed states is more complicated. Though mixed states
in many cases can serve as a QIT resource  it is difficult
to characterise useful mixed states entanglement in general. 
In addition the fundamental question initiated in \cite{Werner}, namely 
which entangled mixed states admitt local hidden variable theories remains 
still open. The very interesting type of entanglement that serves as an 
ideal probe for the above analysis is bound entanglement (BE) \cite{bound} that can not be distilled
to pure entanglend form, nevertheless turns out to be useful in some
quantum QIT tasks \cite{activ,conc,superactivAB,boundkey,super,multiple,Ishizaka}.
On the other hand recently, following pioneering 
and surprising result \cite{Dur}, a few multipartite bound entangled states 
\cite{Kaszlikowski,Acin} including especially the case of 6 qubits 
\cite{Sen}.
Note that this means that BE can serve for reduction of communication complexity in wide class of schemes provided in \cite{complexity1,complexity2}.
The scenario with minimal number of particles $N=6$ required continuous 
setting Bell inequalities that can not be implemented experimentally. 
Also no maximal violation of Bell inequalities have been reported 
for analysed states. 

Quite recently, however, Smolin bound entangled states \cite{Smolin}
representing for qubit density matrices have been reported \cite{my} to 
violate Bell inequalites {\it maximally} in a very simple setting
(similar to CHSH \cite{CHSH} scenario). At the same time 
the states do not admitt multiparty cryptography 
scenario which means that even maximal violation of Bell inequalities
does not imply quantum security if all the parties are in distant labs. 

In the present work we generalise Smolin states to any even number of particles
calling new states generalised Smolin States (GSS).
We show that they maximally violate Bell inequalities as it was in 
the case of four qubits. As such they can reduce communication 
complexity. Still it can be shown, as in four-qubit case that is spite of 
maximal Bell violation, the states are not useful for quantum
security. On the other hand we show that GSS - like the original 
Smolin states (see \cite{conc}) - allow for remote quantum information 
concentration. Quantum network realising the Smolin states is also 
designed. Finally we discuss the relation of Bell inequality violation 
and quantum security. We find a possibility of interesting application 
of the result of information concentration as an unlocking of 
large amount of classical information.

\section{Generalized Smolin states}
\label{II}
\subsection{Construction}
In this section we extend the last developments concerning bound
entanglement in context of Bell inequalities \cite{my} to the case
of arbitrary even number of particles. 

In the very beginning let us define the following class of unitary
operations
\begin{equation}\label{Unitary}
U_{i}^{(n)}=I^{\otimes n-1}\otimes\sigma_{i},\qquad
n=1,2,3,\ldots,\qquad i=0,1,2,3,
\end{equation}
where $\sigma_{0}=I$ is identity acting on $\mathbb{C}^{2}$ and
$\sigma_{i},\;i=1,2,3$ are the standard Pauli matrices. Then let
us introduce the so-called Bell basis defined on Hilbert space
$\mathbb{C}^{2}\ot\mathbb{C}^{2}$ as
\begin{eqnarray}\label{BellStates}
&&\ket{\psi_{0}^{B}}\equiv\ket{\psi_{-}}=(1/\sqrt{2})\,(\ket{01}-\ket{10}),\nonumber\\
&&\ket{\psi_{1}^{B}}\equiv\ket{\phi_{-}}=(1/\sqrt{2})\,(\ket{00}-\ket{11}),\nonumber\\
&&\ket{\psi_{2}^{B}}\equiv\ket{\phi_{+}}=(1/\sqrt{2})\,(\ket{00}+\ket{11}),\nonumber\\
&&\ket{\psi_{3}^{B}}\equiv\ket{\psi_{+}}=(1/\sqrt{2})\,(\ket{01}+\ket{10}).
\end{eqnarray}
From (\ref{Unitary}) and (\ref{BellStates}) one can immediately
infer that
$U^{(2)}_{i}\proj{\psi_{0}^{B}}U^{(2)}_{i}=\proj{\psi_{i}^{B}}$.
For the purposes of further analyzes it is convenient to rewrite
the above states using the Hilbert--Schmidt formalism (see 
\cite{Horodeccy1}. Let us
recall that every state $\varrho$ acting on space
$\mathbb{C}^{2}\otimes \mathbb{C}^{2}$ may be written in the form
\begin{equation}
\varrho=\frac{1}{4} \left( I\otimes
I+\boldsymbol{r}\cdot\boldsymbol{\sigma}\otimes
I+I\otimes\boldsymbol{s}\cdot\boldsymbol{\sigma}+\sum_{i,j=1}^{3}t_{ij}\sigma_{i}\otimes\sigma_{j}
\right),
\end{equation}
where $I$ is defined as previous, $\boldsymbol{r}$ and
$\boldsymbol{s}$ are vectors from $\mathbb{R}^{3}$,
$\boldsymbol{\sigma}$ is vector constructed from Pauli matrices,
i.e., $\boldsymbol{\sigma}=[\sigma_{1},\sigma_{2},\sigma_{3}]$.
Finally, coefficients
$t_{ij}=\mathrm{Tr}(\mathrm{\varrho\sigma_{i}\otimes\sigma_{j}})$
form real-valued matrix $t$. 
For Bell states (\ref{BellStates}) we have 
the nice geometrical structure\cite{Horodeccy1} that 
results in particular in:
\begin{eqnarray}\label{BellinHS}
&&\proj{\psi_{l}^{B}}=\frac{1}{4}\left[I^{\otimes
2}+\sum_{i=1}^{3}t_{ii}^{(l)}\sigma_{i}^{\otimes 2}\right],\qquad
l=1,\ldots,4,\nonumber\\
&& t^{(0)}=\mathrm{diag}[-1,-1,-1]\nonumber\\
&& t^{(1)}=\mathrm{diag}[-1,1,1]\nonumber\\
&& t^{(2)}=\mathrm{diag}[1,-1,1]\nonumber\\
&& t^{(3)}=\mathrm{diag}[1,1,-1].
\end{eqnarray}
Note that for all Bell states vectors $\boldsymbol{r}$ and
$\boldsymbol{s}$ equal to zero and matrices $t^{(i)}$ are
diagonal. Moreover all these states maximally violate CHSH-Bell
inequality for correlation function \cite{CHSH}, i.e., the amount
of violation is $\sqrt{2}$ and it is maximal value achievable by
Quantum Mechanics.

Then let us introduce the so-called Smolin state \cite{Smolin}
acting on space $(\mathbb{C}^{2})^{\ot 4}$:
\begin{eqnarray}\label{Smolin}
\rho^{S}={\textstyle\frac{1}{4}}\sum_{i=0}^{3}\proj{\psi_{i}^{B}}^{\ot
2}={\textstyle\frac{1}{4}}\sum_{i=0}^{3}\left(U^{(2)}_{i}\proj{\psi_{0}^{B}}U^{(2)}_{i}\right)^{\otimes
2},\nonumber\\
\end{eqnarray}
This state is bound entangled since we cannot distill singlet
between any pair of particles. However, the distillation is
possible when any two particles are in the same laboratory. As
shown in \cite{my} the Smolin state posses the intriguing feature,
namely despite being bound entangled it violates maximally the
CHSH-type Bell inequality for four particles.

Now, we are in position to present our method. Firstly let us
define states by the recursive formulas:
\begin{eqnarray}\label{Construction}
&&\rho_{2}\equiv\proj{\psi^{B}_{0}},\nonumber\\
&&\rho_{4}={\textstyle\frac{1}{4}}\sum_{i}U_{i}^{(2)}\rho_{2}U_{i}^{(2)}\ot
U_{i}^{(2)}\rho_{2}U_{i}^{(2)}\equiv\rho^{S},\nonumber\\
&&\rho_{6}={\textstyle\frac{1}{4}}\sum_{i}U_{i}^{(4)}\rho_{4}U_{i}^{(4)}\ot
U_{i}^{(2)}\rho_{2}U_{i}^{(2)},\nonumber\\
&&\vdots\nonumber\\
&&\rho_{2(n+1)}={\textstyle\frac{1}{4}}\sum_{i}U_{i}^{(2n)}\rho_{2n}U_{i}^{(2n)}\ot
U_{i}^{(2)}\rho_{2}U_{i}^{(2)}.
\end{eqnarray}
This construction starts from one of the Bell states, namely,
singlet. Obviously this state is {\it free} entangled state
and violates maximally  Bell inequalities. 
This property is crucial for our purposes
since, as we will see below, our construction is 'smuggling' it to
the arbitrary even number of particles. Furthermore, as it is
underlined in (\ref{Construction}), $\rho_{4}$ is bound entangled
and therefore, again because of this specific type of
construction, all states from this class with $n\ge 4$ are bound
entangled. It is interesting that from all these states with $n\ge
4$ it is possible to distill only one singlet whenever any subset
of $n-2$ particles are in the same laboratory.

Hereafter states $\rho_{2n}$ for $\; n > 2$ we shall call
{\it Generalized Smolin States} (GSS).

It is worth noticing that all these states, including $\rho_{2}$
are permutationally invariant since we have the following

\Obs{1}{ Any state $\rho_{2n}$ may be written in the form}
\begin{equation}\label{Observation1}
\rho_{2n}=\frac{1}{2^{2n}}\left[I^{\otimes
2n}+(-1)^{n}\sum_{i=1}^{3}\sigma_{i}^{\otimes 2n}\right],\qquad
n=1,2,3,\ldots
\end{equation}
\Proof The proof will be established using mathematical induction.
For $n=1$ this observation is obvious, since (\ref{Observation1})
represents Hilbert-Schmidt expression for singlet
(\ref{BellinHS}). Therefore for further clarity we investigate the
case with $n=2$, i.e., the case of Smolin state. So, our task is
to prove that
\begin{equation}\label{}
\rho_{4}\equiv\rho^{S}=\frac{1}{16}\left(I^{\otimes
4}+\sum_{i=1}^{3}\sigma_{i}^{\otimes 4}\right).
\end{equation}
To this aim it suffice to substitute Hilber-Schmidt expansions for
all Bell states to (\ref{Smolin}) and to utilize the facts that
\begin{equation}\label{}
\sum_{i=0}^{3}t^{(i)}=\mathsf{0},\qquad \sum_{i=0}^{3}t^{(i)}\ot
t^{(i)}=4\mathrm{diag}[1,\ldots,1,\ldots,1].
\end{equation}

Now we assume that for arbitrary natural number $n$ the thesis
(\ref{Observation1}) if fulfilled. Then we need to prove that
\begin{equation}\label{A}
\rho_{2(n+1)}=\frac{1}{2^{2(n+1)}}\left[I^{\otimes
2(n+1)}+(-1)^{n+1}\sum_{i=1}^{3}\sigma_{i}^{\otimes
2(n+1)}\right].
\end{equation}
Firstly, let us recall that by the definition (\ref{Construction})
state $\rho_{2(n+1)}$ may be constructed as follows
\begin{equation}\label{B}
\rho_{2(n+1)}=\frac{1}{4}\sum_{i=0}^{3}U_{i}^{(2n)}\rho_{2n}U_{i}^{(2n)}\otimes
U_{i}^{(2)}\rho_{2}U_{i}^{(2)}.
\end{equation}
Secondly, let us note that arbitrary density matrix $\xi$
describing $N$ spin one-half particles may be written as
\begin{equation}\label{generalmixState}
\xi=\frac{1}{2^{N}}\sum_{i_{1},\ldots,i_{N}=0}^{3}
\lambda_{i_{1},\ldots,i_{N}}\sigma_{i_{1}}\ot\ldots\ot\sigma_{i_{N}},
\end{equation}
Coefficients $\lambda_{i_{1},\ldots,i_{N}}$ form tensor that we
shall denote by $\Lambda$ and its part responsible for
$i_{1},\ldots,i_{N}=1,2,3$ by $T$. Immediate observation is that
for states $U_{i}^{(2n)}\rho_{2n}U_{i}^{(2n)}$ all coefficients
$\lambda_{i_{1},\ldots,i_{N}}$ are equal to zero except the cases
where $i_{1}=i_{2}=\ldots=i_{N}$. Moreover, it is clear from
(\ref{Observation1}) and by virtue of the equality
$\sigma_{i}\sigma_{j}\sigma_{i}=2\delta_{ij}\sigma_{i}-\sigma_{j}$,
that states $U_{i}^{(2n)}\rho_{2n}U_{i}^{(2n)}$ have tensors $T$
of the form
\begin{eqnarray}\label{T1}
&&T_{2n}^{(0)}=(-1)^{n}\mathrm{diag}[1,\ldots,1,\ldots,1],\nonumber\\
\qquad
&&T_{2n}^{(1)}=(-1)^{n}\mathrm{diag}[1,\ldots,-1,\ldots,-1],\nonumber\\
&&T_{2n}^{(2)}=(-1)^{n}\mathrm{diag}[-1,\ldots,1,\ldots,-1],\nonumber\\
\qquad
&&T_{2n}^{(3)}=(-1)^{n}\mathrm{diag}[-1,\ldots,-1,\ldots,1].\nonumber\\
\end{eqnarray}
Finally one has
\begin{eqnarray}\label{T2}
&&\sum_{i=0}^{3}T_{2n}^{(i)}=\mathsf{0}, \nonumber\\
&&\sum_{i=0}^{3}T_{2n}^{(i)}\otimes
t^{(i)}=4(-1)^{n+1}\mathrm{diag}[1,\ldots,1,\ldots,1].\nonumber\\
\end{eqnarray}
The proof is now completed after substitution of states
$U_{i}^{(2n)}\rho_{2n}U_{i}^{(2n)}$ and Bell states
(\ref{BellStates}) to (\ref{B}) with the aid of (\ref{T1}) and
(\ref{T2}).
\subsection{Violation of Bell inequality}
Quite recently Brukner {\it et al.} \cite{complexity1} showed that
aside from being one of the most important tools in detection of
quantum non-locality, Bell inequalities constitute criterion of
usefulness of the quantum states in reducing communication
complexity. The prove is constructive since for every Bell
inequality and for broad class of quantum protocols they propose a
multi-party communication complexity problem. Quantum protocols
for this problem are more efficient when one uses quantum state
violating that inequality. Therefore the next step of our analyzes
is to show explicitly violation of one chosen inequality by GSS.

To this aim we consider standard scenario in which $j$-th party
$(j=1,2,\ldots,2n)$ can choose between two dichotomic observables
$\hat{O}_{k_{j}}^{(j)},\quad k_{j}=1,2$. Then all parties measure
simultaneously one of arbitrarily chosen observable. After many
runs of experiment, trying to prove that there does not exist any
LHV model for a given state, they must show violation of arbitrary
Bell inequality. For our purposes it suffice to consider CHSH-type
Bell inequality of the form:
\begin{eqnarray}\label{BellInequality}
&&|E(\underset{2n-1}{\underbrace{1,\ldots,1}},1)
+E(\underset{2n-1}{\underbrace{1,\ldots,1}},2)\nonumber\\
&&+E(\underset{2n-1}{\underbrace{2,\ldots,2}},1)
-E(\underset{2n-1}{\underbrace{2,\ldots,2}},2)|\le 2.
\end{eqnarray}
which may be derived from the more general set of Bell
inequalities  \cite{WW}, \cite{Zukowski} or
using the same technique as for two-particle CHSH Bell inequality
\cite{CHSH}. Function $E$ appearing in (\ref{BellInequality}) is
so-called correlation classically defined as an average of 
the measurement outputs taken over many runs of experiment:
\begin{equation}\label{3.2}
E(k_{1},\ldots,k_{N})=\left\langle\prod_{j=1}^{N}
O^{(j)}_{{k_{j}}} \right\rangle_{avg}.
\end{equation}
In quantum regime the definition is 
\begin{equation}
E_{QM}(k_{1},k_{2},\ldots,k_{2n})(\varrho)=
\mathrm{Tr}[\varrho\hat{O}_{k_{1}}^{(1)}\otimes\hat{O}_{k_{2}}^{(2)}\otimes\ldots\otimes\hat{O}_{k_{2n}}^{(2n)}],
\end{equation}
where in case of spin-$\frac{1}{2}$ particles dichotomic
observables are of the form
\begin{equation}\label{Dichotomic}
\hat{O}_{k_{j}}^{(j)}=\hat{\boldsymbol{n}}_{k_{j}}^{(j)}\cdot\boldsymbol{\sigma},\qquad
k_{j}=1,2,\qquad j=1,2,\ldots,2n,
\end{equation}
where $\hat{\boldsymbol{n}}_{k_{j}}^{(j)}$ denote vectors from
$\mathbb{R}^{3}$, obeying
$|\hat{\boldsymbol{n}}_{k_{j}}^{(j)}|=1$. Let us choose the vectors
\begin{eqnarray}\label{kierunki}
&&\boldsymbol{n}_{1}^{(j)}=\hat{x},\qquad
\boldsymbol{n}_{2}^{(j)}=\hat{y},\qquad
j=1,2,\ldots,2n-1,\nonumber\\
&&\boldsymbol{n}_{1}^{(2n)}=\frac{\hat{x}+\hat{y}}{\sqrt{2}},\qquad
\boldsymbol{n}_{2}^{(2n)}=\frac{\hat{x}-\hat{y}}{\sqrt{2}},
\end{eqnarray}
where $\hat{x}$ and $\hat{y}$ stand for unity vectors directed
along, respectively, $OX$ and $OY$ axes. 
The above choice gives the value
\begin{eqnarray}\label{Q}
&&\hspace{-0.4cm}E_{QM}(\underset{2n-1}{\underbrace{1,..,1}},1)(\rho_{2n})+
E_{QM}(\underset{2n-1}{\underbrace{1,..,1}},2)(\rho_{2n})\nonumber\\
&&\hspace{-0.5cm}+E_{QM}(\underset{2n-1}{\underbrace{2,..,2}},1)(\rho_{2n})-
E_{QM}(\underset{2n-1}{\underbrace{2,..,2}},2)(\rho_{2n})=(-1)^{n}2\sqrt{2},\nonumber\\
\end{eqnarray}
that obviously violate Bell inequality (\ref{BellInequality}).
Moreover, this violation is maximal which can be easily shown  by 
Tsirelson bound \cite{Tsirelson} since for this purpose 
in each term  (\ref{Q}) we can combine 
all 2n-1 local operators into one dychotomic operator.

 Concluding, we have just shown that any of states
(\ref{Construction}) violates Bell inequality
(\ref{BellInequality}) {\it maximally}. However, for $n=1$ it is obvious
since for this value of $n$ we have one of the Bell states, for
$n>1$ this violation is surprising in the light of the fact that
all these states are bound entangled.

\subsection{Noisy states}
Trying to generalize the above considerations, we investigate some
of the properties of the GSS in presence of noise. In other words
below we characterize states
\begin{equation}\label{NoisyGSS}
\varrho_{2n}(p)=(1-p)\frac{I^{\otimes
2n}}{2^{2n}}+p\rho_{2n},\qquad 0\le p\le 1
\end{equation}
where $I$ as previous is identity acting on one-qubit space and
bound entangled states $\rho_{2n}$ are defined by
(\ref{Construction}). Below we show that this family of states has
similar separability properties and violate Bell inequality in the
same regime with respect to $p$ as two-qubit Werner states
\cite{Werner}.

In the first step let us observe that by virtue of
(\ref{Observation1}) we may rewrite (\ref{NoisyGSS}) as follows
\begin{equation}\label{}
\varrho_{2n}(p)=\frac{1}{2^{2n}}\left[I^{\otimes
2n}+(-1)^{n}p\sum_{i=1}^{3}\sigma_{i}^{\otimes 2n}\right]
\end{equation}
To investigate separability properties of (\ref{NoisyGSS}) let us
introduce projectors
\begin{equation}\label{ProjP}
P\p_{k}=\frac{1}{2}(I\pm\sigma_{k})
\end{equation}
as corresponding to eigenvectors of $\sigma_{k}$ with eigenvalues
$\pm1$. Then let us consider two-qubit mixed separable states
introduced in \cite{Horodeccy1}:
\begin{equation}\label{}
\varrho_{k}^{(\pm)}=\frac{1}{2}\left[P_{k}^{(+)}\otimes
P_{k}^{(\pm)}+P_{k}^{(-)}\otimes P_{k}^{(\mp)}\right]
\end{equation}
Please notice that these states may be easily generalized to
arbitrary amount of particles. To this aim let us introduce the
following notations:
\begin{eqnarray}\label{statesEta}
&&\eta_{k,1}\p\equiv P_{k}\p,\nonumber\\
&&\eta_{k,2}\p={\textstyle\frac{1}{2}}\left[\eta_{k,1}^{(+)}\otimes
P_{k}\p+\eta_{k,1}^{(-)}\otimes P_{k}\m\right]\equiv\varrho_{k}\p,\nonumber\\
&&\eta_{k,3}\p={\textstyle\frac{1}{2}}\left[\eta_{k,2}^{(+)}\otimes
P_{k}\p+\eta_{k,2}^{(-)}\otimes P_{k}\m\right],\nonumber\\
&&\vdots\nonumber\\
&&\eta_{k,n}\p={\textstyle\frac{1}{2}}\left[\eta_{k,n-1}^{(+)}\otimes
P_{k}\p+\eta_{k,n-1}^{(-)}\otimes P_{k}\m\right].
\end{eqnarray}
From that construction it is obvious that all states
$\eta_{k,n}\p$ are fully separable. Moreover, taking into account
expression (\ref{ProjP}) we may constitute the following

\Obs{3}{All states $\eta_{k,n}^{(\pm)}$ have the form}
\begin{equation}\label{Observation3}
\eta_{k,n}\p=\frac{1}{2^{n}} \left(I^{\otimes
n}\pm\sigma_{k}^{\otimes n}\right).
\end{equation}
\Proof Since the above observation is rather obvious, we decided
to present below proof for $n=2$. Generalization to arbitrary $n$
is straightforward. From the definition (\ref{statesEta}) we infer
\begin{equation}
\eta_{k,2}\p\equiv\varrho_{k}\p=\frac{1}{2}\left[P_{k}^{(+)}\otimes
P_{k}\p+P_{k}^{(-)}\otimes P_{k}\m\right],
\end{equation}
and then application of (\ref{ProjP}) to the above yields
\begin{eqnarray}\label{}
\eta_{k,2}\p&=&\frac{1}{8}\left[(I+\sigma_{k})\otimes(I\pm
\sigma_{k})+(I-\sigma_{k})\otimes
(I\mp\sigma_{k})\right]\nonumber\\
&=&\frac{1}{2^{2}}\left[I^{\otimes 2}\pm \sigma_{k}^{\otimes
2}\right].\hspace{3cm}\square
\end{eqnarray}
%
%
%
%
%
Now we are in position to finish considerations respecting
separability properties of (\ref{NoisyGSS}). Since the Werner
state $\varrho^{W}(p)$ and the Smolin state $\varrho^{S}(p)$ are
separable for $p=1/3$ we may conjecture that all GSS for $n>2$ are
also separable for such value of $p$. Indeed, we have

\Obs{4}{For $p=\textstyle{\frac{1}{3}}$ states
$\varrho_{2n}(\textstyle{\frac{1}{3}})$ are separable and are of
the form}
\begin{equation}\label{Observation4}
\varrho_{2n}({\textstyle\frac{1}{3}})={\textstyle\frac{1}{6}}\sum_{k=1}^{3}\left\{
\begin{array}{l}
\eta_{k,n}^{(+)}\otimes\eta_{k,n}^{(-)}+
\eta_{k,n}^{(-)}\otimes\eta_{k,n}^{(+)},\\
\hspace{3cm}\quad n=1,3,5,\ldots\\
\eta_{k,n}^{(-)}\otimes\eta_{k,n}^{(-)}+
\eta_{k,n}^{(+)}\otimes\eta_{k,n}^{(+)},\\
\hspace{3cm}\quad n=2,4,6,\ldots\,.
\end{array}
\right.
\end{equation}
\Proof The proof is rather technical, so we restrict our
considerations to the case of odd number of particles. After
application of (\ref{Observation3}) to (\ref{Observation4}) we
obtain
\begin{eqnarray}\label{}
&&\hspace{-1cm}\varrho_{2n}({\textstyle\frac{1}{3}})=\frac{1}{6}\frac{1}{2^{2n}}\sum_{k=1}^{3}
\left[ \left( I^{\ot n}+\sigma_{k}^{\ot n} \right)\ot\left( I^{\ot
n}-\sigma_{k}^{\ot n} \right)\right.\nonumber\\
&&\hspace{-0.8cm}+\left.\left( I^{\ot n}-\sigma_{k}^{\ot n}
\right)\ot\left( I^{\ot n}+\sigma_{k}^{\ot n}
\right)\right]\nonumber\\
&&\hspace{-0.8cm}=
\frac{1}{2^{2n}}\left(I^{\ot
2n}-\frac{1}{3}\sum_{k=1}^{3}\sigma_{k}^{\ot 2n}\right),\qquad
n=1,3,5,\ldots
\end{eqnarray}
The same procedure for the even number of particles gives
expression with minus before the sum. 
Thus, rewriting these two relations in generalized form
%
%
\begin{equation}\label{}
\varrho_{2n}({\textstyle\frac{1}{3}})=\frac{1}{2^{2n}}\left[I^{\ot
 2n}+(-1)^{n}\frac{1}{3}\sum_{k=1}^{3}\sigma_{k}^{\ot 2n}\right],
\end{equation}
completes the proof. Remark that using LOCC we may always add some
noise and therefore noisy GSS become separable for all $p\in
[0,1/3]$. Subsequently, using observables defined by
(\ref{kierunki}), we can see that violation of
(\ref{BellInequality}) by (\ref{NoisyGSS}) is for $p\in
(1/\sqrt{2},1].$
\section{Applications}

\subsection{Communication Complexity}
It is quite remarkable that \cite{my} despite  being bound entangled
Smolin states can reduce communication complexity. This fact shows that,
however, bound entangled states are not distillable, they allow to
solute some tasks with the same efficiency like free entangled
states. Here we show that the Smolin state is not an isolated
case, where bound entangled states are equal to free entangled
states in context of reducing communication complexity. As proven
by \cite{complexity1} the necessary and sufficient condition for
being useful in reducing communication complexity is violation of
one arbitrary Bell inequalities. In the light of the former we can
see that all GSS are useful with the same efficiency as free
entangled states, since we have already proven that this violation
is maximal.

\subsection{Remote information concentration}
Before we prove the utility of GSS in remote information
concentration we focus on telecloning scheme proposed by Murao
{\it et al.} \cite{Telecloning}. This scheme, involving quantum
teleportation and cloning allow a sender to teleport an unknown
qubit state to spatially separated receivers. Of course, in virtue
of no-cloning theorem received qubits are no longer
perfect clones of teleported one. On the other hand it is shown
that fidelities achievable in such a scheme are sufficient.
Suppose that Alice wishes to teleport one qubit $\ket{\phi}_{X}$
to her spatially separated friends $B_{1},\ldots,B_{M}$. After all
(for more details see \cite{Telecloning}) all of them share
so-called {\it optimal cloning state}
\begin{equation}\label{}
\ket{\Psi_{c}}=a\ket{\phi_{0}}_{AC}+b\ket{\phi_{1}}_{AC},
\end{equation}
where
\begin{eqnarray}\label{}
&&\ket{\phi_{0}}_{AC}=\sum_{j=0}^{M-1}\alpha_{j}\ket{A_{j}}_{A}\ot\ket{\{0,M-j\},\{1,j\}}_{C}\nonumber\\
&&\ket{\phi_{1}}_{AC}=\sum_{j=0}^{M-1}\alpha_{j}\ket{A_{M-1-j}}_{A}\ot\ket{\{0,j\},\{1,M-j\}}_{C}\nonumber\\
&&\alpha_{j}=\sqrt{\frac{2(M-j)}{M(M+1)}},
\end{eqnarray}
and
\begin{equation}\label{}
\ket{A_{j}}_{A}=\ket{\{0,M-1-j\},\{1,j\}}_{A}.
\end{equation}
Kets $\ket{A_{j}}_{A}$ represent $M$ normalized and orthogonal
states of ancilla involving $M-1$ qubits. The subscript $C$ refers
to $M$ qubits holding the clones and finally ket
$\ket{\{0,M-j\},\{1,j\}}$ stands for normalized and symmetric
state of $M$ qubits. Let us notice that
\begin{eqnarray}\label{}
&&\underset{2M-1}{\underbrace{\sigma_{3}\ot\ldots\ot\sigma_{3}}}\ket{\phi_{l}}_{AC}
=(-1)^{l}\ket{\phi_{l}}_{AC},\nonumber\\
&&\underset{2M-1}{\underbrace{\sigma_{2}\ot\ldots\ot\sigma_{2}}}\ket{\phi_{l}}_{AC}=
(-1)^{M+l+1}\mathrm{i}\ket{\phi_{l\oplus 1}}_{AC},\nonumber\\
&&\underset{2M-1}{\underbrace{\sigma_{1}\ot\ldots\ot\sigma_{1}}}\ket{\phi_{l}}_{AC}=\ket{\phi_{l\oplus
1}}_{AC},
\qquad l=0,1,\nonumber\\
\end{eqnarray}
where $\oplus\equiv +_{\mathrm{mod2}}$.

Murao and Vedral proved \cite{conc} that even if clones are not
perfect replicas of teleported qubit, it is still possible to
recover information included in optimal cloning state to Charlie
using only LOCC. To show it explicitly they used unlockable bound
entangled Smolin state $\rho^{S}$. Now we show that all GSS are
useful to perform such a task.

At the very beginning let us assume that the optimal cloning state
is distributed among Alice and her friends $B_{1},\ldots,B_{M}$ in
such a way that the former posses $M-1$ ancilla qubits (generally
these qubits may be also spatially separated) and the latter $M$
qubits of clones. Subsequently, they wish to recreate the original
qubit to Charlie using as a quantum channel GSS distributed
previously among all actors. To complete this goal Alice and
$B_{1},\ldots,B_{M}$ perform a Bell measurement between their
qubits, one from optimal cloning state, and one from GSS. After
that they arrive at the state $\varrho_{k_{1}\ldots k_{N}}$ given
by
\begin{eqnarray}\label{pomiar}
&&\hspace{-0.5cm}\varrho_{k_{1}\ldots
k_{N}}=\frac{1}{p_{k_{1}\ldots
k_{N}}}\nonumber\\
&&\times\mathrm{Tr}_{A_{1},..,A_{M-1},B_{1},..,B_{M}}
\left[\left(\bigotimes_{i=1}^{N}P_{k_{i}}\ot U_{k_{1}k_{2}\ldots
k_{N}}\right)\right.\nonumber\\
&&\left.\proj{\Psi_{c}}\ot
\rho_{2M}\left(\bigotimes_{i=1}^{N}P_{k_{i}}\ot
U^{\dagger}_{k_{1}k_{2}\ldots k_{N}}\right)\right],
\end{eqnarray}
with probability $p_{k_{1}\ldots
k_{N}}=\mathrm{Tr}_{C}(\varrho_{k_{1}\ldots k_{N}})$, where
$N=2M-1$. To compute all traces appearing in (\ref{pomiar}) we
shall utilize the same technique as presented in \cite{Horodeccy2}
(see also \cite{Linden}). Therefore it is convenient to take the
optimal cloning state $\ket{\Psi_{c}}$ in form
(\ref{generalmixState}). Hence, after substitution of
(\ref{Observation1}) to (\ref{pomiar}) we have
\begin{eqnarray}\label{pomiar2}
&&\hspace{-1cm}\varrho_{k_{1}\ldots k_{N}}=\frac{1}{p_{k_{1}\ldots
k_{N}}}\frac{1}{2^{2M+N}}\sum_{m_{1},\ldots,m_{N}}\lambda_{m_{1}\ldots
m_{N}}\nonumber\\
&&\hspace{-0.8cm}\times\left\{\prod_{l=1}^{N}\mathrm{Tr}[P_{k_{l}}(\sigma_{m_{l}}\ot
I)]I+(-1)^{M}\right.\nonumber\\
&&\hspace{-0.8cm}\times\left.\sum_{r=1}^{3}\prod_{l=1}^{N}\mathrm{Tr}[P_{k_{l}}(\sigma_{m_{l}}\ot\sigma_{r})]
U_{k_{1}k_{2}\ldots k_{N}}\sigma_{r}U^{\dagger}_{k_{1}k_{2}\ldots
k_{N}}\right\}
\end{eqnarray}
Since
\begin{eqnarray}\label{}
&&\mathrm{Tr}[P_{k_{l}}(\sigma_{m_{l}}\ot
I)]=\delta_{m_{l}0}\nonumber\\
&&\mathrm{Tr}[P_{k_{l}}(\sigma_{m_{l}}\ot \sigma_{r})]=
\sum_{r=1}^{3}t_{ii}^{(k_{l})}\delta_{im_{l}}\delta_{ir}
\end{eqnarray}
we may rewrite (\ref{pomiar2}) as
\begin{eqnarray}\label{pomiar3}
&&\hspace{-1cm}\varrho_{k_{1}\ldots k_{N}}=\frac{1}{p_{k_{1}\ldots
k_{N}}}\frac{1}{2^{2M+N}}\Big[\lambda_{0\ldots 0}I\nonumber\\
&&\hspace{-0.5cm}+(-1)^{M}U_{k_{1}\ldots k_{N}}(t^{(k_{1})}\ldots
t^{(k_{N})}\vec{\lambda})\cdot\vec{\sigma}U^{\dagger}_{k_{1}\ldots
k_{N}}\Big],
\end{eqnarray}
with $\vec{\lambda}=[\lambda_{1\ldots 1},\lambda_{2\ldots
2},\lambda_{3\ldots 3}]$. As we shall see below it suffice for
Charlie to perform an operation\linebreak $U_{k_{1}\ldots
k_{N}}=\sigma_{2}^{M\oplus 1}\sigma_{k_{1}}\ldots \sigma_{k_{N}}$
in order to obtain an original qubit. Since
\begin{equation}\label{}
\sum_{k=0}^{3}\sigma_{k}(t^{(k)}\vec{\lambda})\cdot\vec{\sigma}\sigma_{k}=-4\vec{\lambda}\cdot\vec{\sigma}
\end{equation}
we finally obtain
\begin{eqnarray}\label{}
&&\hspace{-1cm}\sum_{k_{1},\ldots,k_{N}=0}^{3}p_{k_{1}\ldots
k_{N}}\varrho_{k_{1}\ldots k_{N}}\nonumber\\
&&=\frac{1}{2}[\lambda_{0\ldots 0}I+(-1)^{M+1}\sigma_{2}^{M\oplus
1}\vec{\lambda}\cdot\vec{\sigma}\sigma_{2}^{M\oplus 1}]
\end{eqnarray}

\section{Discussion}

Let us consider some of the properties of generalised Smolin states.
It is interesting to understand why maximal Bell violation 
does not imply quantum security in this case. The naive approach would 
say that there is some correlations that 
(i) are strictly nonlocal (since all the measurements 
in Bell measurement are performed locally) (ii) are not accessible to Eve
(since Bell inequalities are violated).
One could argue that  any violation singles out one particle 
versus the remaining ones. However still the remaining parties 
perform measurements {\it locally} which means that   
one the correlations are stronger than
just as if they were interpreted in terms 
of entanglement of single particle 
versus all other parties taken together.
On the other hand we have obvious argument against 
security since the states are biseparable (separable against 
any (2)-(2n-2) particles cut) which means 
that no security can be distilled even if some parties 
can communicate quantumly.
Most probably the reason is that the present states, 
despite violating Bell inequalities 
do not have any set of axies that provide perfect correlations 
between all the parties. Thus, in a sense, the quantum 
correlations even if nonlocal are completely useless for 
establishing the correlated data. 

Such a set of axes with corresponding maximally correlated probabilities
is possesed by GHZ state Hilbert-Schmidt representation of which has  
nonvanishig coefficients not only at $\sigma_{i}^{\otimes 2n}$ operators (as GSS havs) but also at all permutations
of opeators $\sigma_{3}^{\otimes 2k} \otimes I^{\otimes 2(n-k)}$.
This allows very easily to design Ekert scheme \cite{Artur} with any of observers 
choosing randomly one of the following four axes: 
$\hat{x}$, $\hat{y}$, $(\hat{x} + \hat{y})/\sqrt{2}$, 
$(\hat{x} - \hat{y})/\sqrt{2}$.
This is not the case in GSS case where only few terms 
survive in Hilbert-Schmidt representation.

In presence of recent important results on security 
in post quantum theories \cite{post}, following the 
above discussion it is reasonable to conjecture that 
any physical system, even in post-quantum theory, that 
maximally violate Bell inequalities leads to cryptographic security if only has 
one pair of axes with maximal correlations.
It would be also interesting to consider the cases when the presence of 
maximal correlations is accompanied nonmaximal 
Bell inequalities violation.


Let us pass to the another interesting issue - remote quantum information  concentration. One can very easily 
to see that it can have application if we apply notion of locking of classical correlations \cite{locking}.
Namely suppose that some huge amount of classical correlations (secure or not) between 2n observers is locked
by single qubit that is further deliberately encoded into many qubits
send to different them. It happens then that GSS allows them to unlock
this classical information in a simple way using 
remote information concentration and further ,,unlocking'' measurement 
of the qubit. It is interesting to note
that in this case quantumness of remotely concentrated information 
is important because of {\it quantum entanglement} of this qubit with
another quantum system that contains the classical locked information.
The above application would be quite powerful if one could rigoriously 
show that it is impossible to unlock the quantum information by 
simulation of the unlocking quantum measurement on the distributed qubit.
The reasoning applies immediately to entanglement locking effect 
\cite{locking1} since in that case one also have to localise qubit in one of distant
labs.

This work is supported by European Union (project RESQ) the Polish Ministry of Scientific Research,
Information Technology (Grant No. PBZ-Min-008/P03/ 03) and 
the Fujitsu Laboratories (Europe). After completing this work we became aware of 
the work on similar subject \cite{BoundEnt}.




\end{document}